\documentclass[%
 reprint,
superscriptaddress,
groupedaddress,
 amsmath,amssymb,
 aps,
pra,
]{revtex4-2}
\usepackage[inkscapelatex=false]{svg}
\usepackage{graphicx}% Include figure files
\usepackage{dcolumn}% Align table columns on decimal point
\usepackage{bm}% bold math
\usepackage{todonotes}

\usepackage{hyperref}% add hypertext capabilities
\usepackage{braket}
\usepackage{subcaption}

\begin{document}

\title{On the Constant Depth Implementation of Pauli Exponentials}
\author{Ioana Moflic}
    \email{ioana.moflic@aalto.fi}
\author{Alexandru Paler}%
 \email{alexandru.paler@aalto.fi}
\affiliation{%
 Aalto University, Espoo, Finland\\
}%

% \begin{abstract}

% \end{abstract}

\maketitle

\section*{Abstract}
We decompose, under the very restrictive linear nearest-neighbour connectivity, $Z^{\otimes n}$ exponentials of arbitrary length into circuits of constant depth using $\mathcal{O}(n)$ ancillae and two-body XX and ZZ interactions. Consequently, a similar method works for arbitrary Pauli exponentials. We prove the correctness of our approach, after introducing novel rewrite rules for circuits which benefit from qubit recycling. The decomposition has a wide variety of applications ranging from the efficient implementation of practical fault-tolerant lattice surgery computations, to expressing arbitrary stabilizer circuits via two-body interactions only and parallel decoding of quantum error-correcting computations.

\section{Introduction}

The efficient compilation of Pauli exponentials has far-reaching implications in quantum computing. Whenever the exponentiation angle is $\pi/2$, Pauli strings can be used as measurement operators for stabilizing quantum systems (e.g.~\cite{roffe2018protecting, grans2024improved, gidney2023pair}). When the angle is part of the set $[\pi/2, \pi/4, \pi/8]$, the exponentiated Pauli strings can be interpreted as multi-body measurements which implement fault-tolerant, lattice surgery computations~\cite{litinski2019game, watkins2024high, cowtan2024ssip}. In the case of arbitrary rotation angles, these strings are central to VQE-algorithms (e.g.~\cite{grimsley2019adaptive}) and to the form of UCCSD terms in quantum chemistry~\cite{barkoutsos2018quantum}. Pauli exponentials are phased gadgets in the ZX-calculus~\cite{cowtan2019phase}.

We introduce a novel decomposition of Pauli exponentials, which uses a linear number of ancillas (effectively doubling the number of qubits) and only two-body XX and ZZ interactions. Depending on where the Pauli exponentials are executed, the \emph{interactions} refer to either native gates or logical operations implemented by projective measurements. As native gates these are easy to implement and engineer in Majorana computers~\cite{grans2024improved}, ion traps~\cite{Debnath_2016}, silicon spin qubits~\cite{ustun2024single}. As logical operations, the interactions are implemented, for example, by lattice surgery on topological QEC (e.g. surface code). In other words, we present a very efficient way of emulating the all-to-all connectivity (on linear nearest-neighbour architectures -- error-corrected or not) required for executing exponentiated Pauli strings.

Increased interest is being expressed in finding time-optimal algorithms for calculating Pauli decompositions of arbitrary complex matrices~\cite{georges2024paulidecompositionfastwalshhadamard}
and in designing optimal compilers for arbitrary Pauli operators~\cite{smith2024optimallygeneratingmathfraksu2nusing}. Our method complements these efforts, by providing a way to efficiently execute any Pauli string exponential via two-body XX and ZZ interactions, even under the constraints of nearest-neighbour topologies. To the best of our knowledge, our constant depth implementation is the most efficient known decomposition of such strings, and has the advantage of significantly lowering the cost of implementing arbitrary computations.

\begin{figure}[!t]
    \centering
    \includegraphics[width=1.0\columnwidth]{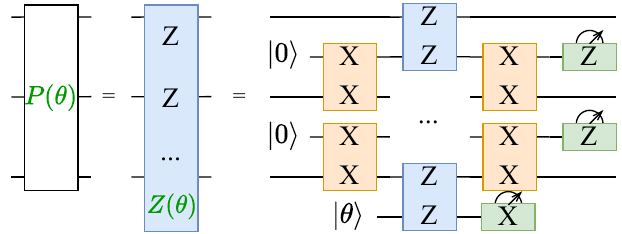}
    \caption{ An $n$-qubit Pauli exponential is decomposed into a constant depth sequence of XX and ZZ interactions and $\mathcal{O}(n)$ ancillae. We use the shorthand notation $e^{-i \theta P}=P(\theta)$, and consider that $P$ is formed exclusively of $Z$ terms: $P = Z^{\otimes n}$. A general string can always be decomposed into single-qubit Clifford gates conjugating the $Z$-terms of $P$~\cite{cowtan2019phase}. In this figure, we assume that $\theta$ is arbitrary, such that our construction is extending $P$ with an additional $Z$ term (marked green) and an ancilla initialized in $\ket{\theta}=R_z(\theta)\ket{+}$~\cite{litinski2019game}.}
    \label{fig:decomp}
\end{figure}

In the following, we discuss the novelty of our method and introduce the building blocks of our derivation in the Methods section. We also briefly describe the wide applicability of our decomposition in the Applications section. The correctness of our method, as well as the complete proof of the decomposition from Fig~\ref{fig:decomp} can be found in Figs~\ref{fig:appb},~\ref{fig:appc},~\ref{fig:appa},

% \subsection{Novelty}

In this work: 1) we offer exact numbers (in contrast to~\cite{yang2024harnessing, buhrman2024state}) for the depth of general Pauli exponentials; 2) introduce a new set of circuit rewrite rules; 3) present the implementation of our method for fault-tolerant computers, computers with restricted topologies as well as qudit computers; 4) discuss the implications on fault-tolerant decoding, offering a novel path towards the approach presented by~\cite{bombin2023modular}.

The constant-depth execution of Pauli exponentials was previously achievable, however, emulating all-to-all connectivity incurred substantial costs. For NISQ machines the costs are, for example, either quadratic qubit overheads and complex four-body interactions~\cite{lechner2015quantum}, the shuttling time in ion trap and neutral atom computers, or the number of SWAP gates in superconducing chips. In error-corrected machines, the costs arise from the difficulty of engineering the surface code buses, which facilitate long-range interactions (e.g.~\cite{saadatmand2024fault}). Other costs come in the form of resource intensive layouts of logical qubits (e.g.~\cite{litinski2019game, beverland2022surface}), or even complex decoding strategies (e.g.~\cite{cain2024correlated}).

Without being generalized to arbitrary rotation angles or number of qubits, constructions similar to ours appeared in~\cite{grans2024improved, gidney2023pair}. Those constructions were derived and verified by applying the ZX-calculus on smaller-scale diagrams. In contrast, we introduce novel circuit rewrite rules, derive and show the correctness of our decomposition using circuit diagrams. We could have used ZX~\cite{de2022circuit} for illustrating the correctness of the derivation from Fig.~\ref{fig:appb} through spider fusions and strong complementarity, however we discovered novel circuit identities (Figs.~\ref{fig:mr},~\ref{fig:fuse}) through circuit rewrites, thus it felt more natural to us to stay in the realm of quantum circuits.

\section{Results}
\label{sec:apps}

Two-body interactions such as XX and ZZ are commonly used in a wide variety of quantum computations and architectures. Herein, we list a few of the applications enabled by our novel method.

\subsection{Lattice Surgery}

\begin{figure}[!h]
    \centering
    \includegraphics[width=1\columnwidth]{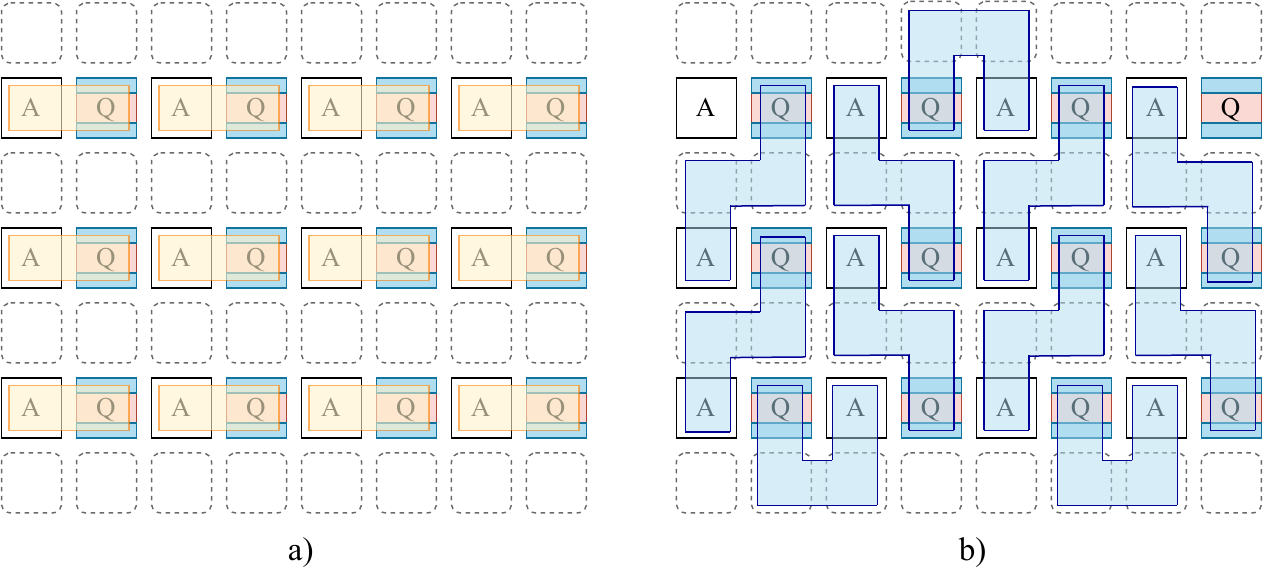}
    \caption{Multi-body $Z^{\otimes n}$ interaction implemented by lattice surgery. The patch layout is adapted from ~\cite{lao2018mapping}: pink patches (Q) hold logical qubits, and gray patches (A) are ancilla for XX and ZZ interactions. The blue boundaries are the Z operators of the logical A patches. a) according to Fig.~\ref{fig:decomp}, we initialize twelve ancilla A in $\ket{0}$ and perform XX two-body interactions (orange) with the Q patches; b) afterwards we perform pairwise ZZ two-body interactions (blue) between pairs of Q and A patches. Finally, the interactions from a) are repeated. The interaction schedule from a) and b) is highly regular and can be further optimized in terms of hardware overheads.}
    \label{fig:lsl}
\end{figure}

Lattice surgery is the de facto standard way of implementing computations with QLDPC codes~\cite{cowtan2024ssip}, such as the surface code. Therein, the logical qubits of the circuits are arranged on a 2D layout of patches (e.g. Fig.~\ref{fig:lsl}), and ancillary space between the patches supports long range multi-body measurements. The latter are expressed as Pauli exponentials. The structure of the layout plays a significant role in the trade-off between a computation's speed and the required amount of physical hardware~\cite{lao2018mapping, litinski2019game}. Although our decomposition uses ancillae, these are readily available in most of the lattice surgery layouts considered for compiling fault-tolerant computations.

The general compilation strategy for lattice surgery is to first decompose arbitrary algorithms into Pauli exponentials~\cite{litinski2019game}, and then to implement these exponentials via multi-body measurements. This approach has multiple disadvantages.

The resulting Pauli exponentials become very long-range and cover large portions of the layout. This hinders the parallelisation of the operations, and alternative compilation methods have been proposed to overcome this, such as~\cite{beverland2022surface} (shows how to perform a constant depth CNOT using two-body measurements), and~\cite{leblond2023realistic}, which compiles to Clifford+T instead of exponentiated Paulis.

Our approach is directly applicable to the compilation method from~\cite{litinski2019game} -- we do not require compiling to CNOTs and then decomposing them to irregular XX and ZZ measurement patterns. The advantage of our decomposition is the very regular measurement schedule (Fig.~\ref{fig:lsl}), which might enable high degrees of parallelism and massive optimizations when considering \textbf{FUSE}.

\subsection{Parallel Decoding}

Our decomposition enables the parallelisation of decoding by setting an upper bound on the maximum (equivalent) decodable distance that a QEC decoder will need to handle within one QEC cycle. In the following, we introduce the concept of equivalent decodable distance.

An additional disadvantage of long-range Pauli exponentials is the large number of physical qubits (data and syndrome) that need to be decoded. For distance $d$, a surface code patch will require $2d^2 - 1$ physical qubits (data and syndrome). The decoding time scales with the size of the surface code patch. Usually, the decoding time is upper bounded by a low-degree polynomial, however, in practice it is still challenging to decode distances larger than 21 within the cycle time of the surface code.

In general, the lattice surgery merge operation between two surface code patches will cover $q=(m+1)(2d^2 - 1)$ qubits, where $m$ is the Manhattan distance between the patches. Decoding the long-range interaction is equivalent to decoding a single patch of distance approximately $\sqrt{\frac{q}{2(m+1)}}$.

As an example, merging two neighboring patches of distance 21 will yield a patch covering $2\times 881 = 1762$ qubits. This is roughly equivalent to a distance 29 patch. If the patches are separated by a single logical patch of distance 21, the equivalent decoding distance would be 36.

In practice, decoding is very challenging. The computational complexity of the decoding algorithms scales polynomially (more than linear)~\cite{demarti2024decoding}, and surface code compilers do not guarantee short-range implementations of Pauli exponentials. Assuming a logical computation of 100 logical qubits, a compact, resource efficient layout of surface code patches would consist of approximately 150 patches~\cite{litinski2019game}. The average Manhattan distance between two arbitrary patches would be around 25, and the average equivalent decoding distance would be 105. For an instance of Shor's algorithm with 2048 logical qubits, the best-case equivalent decoding distance would be around 200. For a decoder with cubic runtime complexity~\cite{demarti2024decoding}, this would have an $\approx (10^2)^3=10^6$ increase in decoding time compared to decoding distance 21.

Our decomposition sets an upper bound on the equivalent decoding distance of arbitrary Pauli exponentials implemented via lattice surgery. For example, in Fig.~\ref{fig:lsl} the blue interactions always cover only four patches. Assuming the distance of the patches is 21, the maximum equivalent decoding distance to decode would be 42.

At the same time, as seen in Fig.~\ref{fig:lsl}, all blue patches can be decoded in parallel. Our parallelisation method is reminiscent of the proposal from~\cite{fowler2013minimum} (fixing the number of physical qubits per decoder) and the constructions from~\cite{poulin2006optimal, skoric2023parallel} (decoders which communicate messages). For the latter, the decoders will communicate the observed correlated logical errors~\cite{cain2024correlated} after the long-range interaction is split via lattice surgery. Our method offers the same decoding benefits like~\cite{bombin2023modular}, with the caveat that we generate automatically a regular pattern of logical blocks, our the size of our blocks is upper bounded.

We conjecture that our decomposition, together with the methods from~\cite{fowler2013minimum, poulin2006optimal, cain2024correlated} will lead to the same decoding performance for Pauli exponentials of arbitrary length.

\subsection{Constant Depth Fault-Tolerant Protocols}

Recently, there has been an increased interest in using only two-body interactions to implement error correction (e.g.~\cite{gidney2023pair, grans2024improved}). Two-body interactions are preferred as quantum computers have limited connectivity between the qubits, and very often the connectivity is not long range (e.g.~\cite{baumer2024measurement}). 

From the perspective of stabilizer QECC codes, our decomposition allows any set of QECC stabilizers to be implemented in constant depth using pairwise XX and ZZ measurements. This claim follows for Z-strings from  Fig.~\ref{fig:decomp} and~\ref{fig:pg}) and requires two additional depth-one layers of single qubit Clifford gates (for a total depth-five construction) to achieve arbitrary stabilizers.

QECC code conversion schemes~\cite{anderson2014fault} can also be implemented locally using pairwise interactions. Additionally, after carefully choosing the pairwise interaction schedules, our method might lead to results similar to the ones from~\cite{alam2024dynamical}. Our decomposition involves only two-body measurements and may find relevance to subsystem codes such as the Bacon-Shor code~\cite{bacon2006operator} or its Floquet version~\cite{alam2024dynamical}.

Our method is also useful for the implementation of Clifford circuits in constant depth by using only pairwise interactions. For example, the authors of~\cite{zheng2018depth} show a way of implementing Clifford circuits fault-tolerantly in constant depth via non-local, multi-qubit Pauli operator measurements. We achieve a local implementation of the circuits at the cost of doubling the number of qubits.

It should be noted that the listed constructions might not be fully fault-tolerant, e.g. introduce hook errors in certain scenarios, and such analysis is left for future work.

\subsection{NISQ computations}

The recursive decomposition of Pauli exponentials (i.e. phase gadgets) using \textbf{PG} will result in a ladder of CNOTs. These ladders are frequently encountered in the circuits expressing UCCSD terms. The long range CNOT construction from~\cite{baumer2024measurement} has already recognized that such terms can be implemented in constant depth.

Our decomposition achieves the same constant depth, but this time using two-body interactions which are practically native gates at the hardware level. In conjunction with architectures such as~\cite{ustun2024single}, our decomposition can offer speed-ups to the execution of adaptive VQE~\cite{grimsley2019adaptive} or other QAOA-like algorithms.

\section{Discussion} 

We introduced a novel and very space-time efficient decomposition of Pauli exponentials. Compared to previous methods of implementing long range interactions (e.g. LHZ~\cite{lechner2015quantum}) our method uses only local two-body interactions, has a linear qubit cost (versus quadratic qubit costs, or time intensive shuttling or SWAP gate schedules) and is applicable to a very wide range of quantum computations, from NISQ to the automatic Floquet-itification of QEC codes. 

Our decomposition is applicable to both NISQ and error-corrected machines. For the latter, we are still assuming that each logical operation is implemented with $d$ rounds of error correction, and the constant-depth refers to the computation's space-time volume~\cite{saadatmand2024fault}, which is agnostic of the code's distance. This does not exclude the possibility of implementing Clifford logical operations in $\mathcal{1}$ rounds of error correction~\cite{cain2024correlated} based on improved decoding (which our method enables, but is left for future work).

A result concerning long range CNOTs has been proposed by~\cite{baumer2024measurement}. Their adaptive circuit construction uses the same number of ancillae as our decomposition. Nevertheless, due to the exclusive focus on long-range CNOT gates, it does not solve the issue of the Clifford-ladder~\cite{yang2024harnessing,cowtan2019phase}. Another constant-depth implementation of long range CNOTs using pairwise measurements is presented in~\cite{beverland2022surface}, and is motivated by the fact that efficient decompositions of Pauli exponentials were not known at the time. Our work bridges the gap that motivated~\cite{beverland2022surface}.

Our method is complementary to the works of~\cite{yang2024harnessing, beverland2022surface, buhrman2024state}. Therein, the focus was the standard quantum circuit view of computation based on Clifford gates (e.g. CNOTs) and classical measurements, in order to implement low-depth computations using Bell pairs and more general GHZ states. However, we are not using the \emph{trick} of decomposing~\cite{yang2024harnessing, beverland2022surface} or deforming circuits~\cite{buhrman2024state} into an entangling part followed by a computation/gate part, and a classical processing part -- and approach which is similar to works such as~\cite{paler2017fault, vijayan2024compilation, kaldenbach2024mapping}. In contrast, we take a different approach and treat Pauli strings as a native gate set, specific for some quantum computing architectures and general fault-tolerant computations.

% Our approach is also enabling the parallelisation and message passing distribution of QEC decoders. Future work will focus on a more detailed analysis of mentioned applications.

\section{Methods}
\label{sec:methods}

\begin{figure}[!t]
    \centering
    \includegraphics[width=0.5\columnwidth]{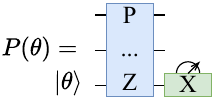}
    \caption{The \textbf{ROT} rewrite rule implements the exponential of the Pauli $P$ by introducing an ancilla initialized in $\ket{\theta}=R_z(\theta)\ket{+}$, extending the Pauli string (the blue box) with a $Z$-term for the ancilla, and measuring the ancilla in the X-basis~\cite{litinski2019game}.}
    \label{fig:litinski}
\end{figure}

The string $P$ can be exponentiated to a unitary by computing $e^{-i \theta P}$. We consider the measurement-based implementation of the exponentiated Pauli strings~\cite{litinski2019game}, such  that, in practice, the implementation uses \emph{an additional Z term in the Pauli string} (green in Fig.~\ref{fig:decomp}) and an ancilla rotated by $\theta$. Consequently,  $e^{-i \theta P}$ is achieved by the rule \textbf{ROT} from Fig.~\ref{fig:litinski}.

The lhs. of Fig.~\ref{fig:decomp} represents the unitary obtained by the exponentiation. The rhs. is the decomposition which includes ancillae for enabling the XX interactions, as well as the $\ket{\theta}$ ancilla initialized for performing a teleportation-based gate. In all the following figures, blue boxes represent Pauli strings that include Z terms, and orange boxes are Pauli strings consisting entirely of X.

\subsection{Measurement-based implementation}
\label{sec:mb}

When compiling a computation into Pauli exponentials on fault-tolerant machines, the $\theta$ rotation has to be decomposed beforehand into a sequence of rotation gates supported by the QEC~\cite{saadatmand2024fault}. This increases the computation's depth proportionally to the length of the gate sequence approximating $\theta$. Moreover, the depth will additionally increase by a constant factor, as each rotation is probabilistic and might require corrections~\cite{litinski2019game} (e.g. if the X-measurement results in the -1 eigenvalue, then a corrective rotation of $2\theta$ is needed). In the following we assume that the fault-tolerant computation has already been compiled, and when discussing the implementaiton of a logical Pauli exponential we refer to the circuit that is consuming the (approximated) rotated state $\ket{\theta}$ (the lowest qubit in Fig.~\ref{fig:decomp}), and show that this circuit has a constant depth with respect to two-body XX/ZZ interactions.

Moreover, in the fault-tolerant case, we also assume that the two-body interactions are measurement-based. The latter will introduce corrective terms (e.g.~\cite{gidney2023pair}), but the corrections can be tracked in a Pauli frame, as the underlying circuit will consist solely of CNOT gates and ancillae initialised or measured in a discrete set of rotated basis~\cite{paler2017fault, vijayan2024compilation}.

If the exponentials are implemented on NISQ machines, the $\theta$ rotation is a single qubit gate. The corrective terms generated by the measurements of the ancillae will be applied adaptively without increasing the depth of the circuit~\cite{baumer2024measurement}.

\emph{As a conclusion, in all the following diagrams we omit the corrective terms introduced by the measurements. The diagrams with corrections included can be found in Supplementary Material. Corrections can be tracked through the circuit, or can be applied immediately after the exponentiated Pauli without affecting the depth of the decomposition, as shown in the Supplementary Material}.

\subsection{Decomposing the Pauli exponentials}

We use the graphical calculus of quantum circuits for reasoning about the decomposition of the Pauli exponentials. To this end, we first present the rewrite rules used to obtain the generalized decomposition from Fig.~\ref{fig:decomp}. The result from Fig.~\ref{fig:decomp} is obtained by applying the following algorithm:
\begin{enumerate}
    \item \emph{Input:} arbitrary Pauli exponential $P$ with angle $\theta$;
    \item Apply \textbf{ROT}, if $\theta$ is not $\pi/2$ -- this extends the Pauli string by an additional Z term and the circuit will include the $\ket{\theta}$ ancilla; if the angle is $\pi/2$ the Pauli string is left unchanged and $\ket{\theta}$ is not appended;
    \item Repeat the sequence of rewrite rules $\textbf{PG, LS, MR, FUSE}$, until the largest Pauli string acts on a maximum of two qubits.
\end{enumerate}

\begin{figure}[!t]
    \centering
    \includegraphics[width=.5\columnwidth]{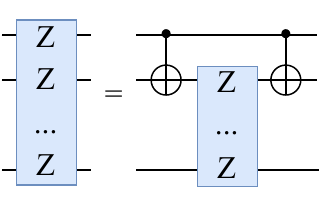}
    \caption{The \textbf{PG} rewrite takes an $n$-qubit string and decomposes it into a $(n-1)$-qubit string and two CNOTs.}
    \label{fig:pg}
\end{figure}

\begin{figure}[!h]
    \centering
    \includegraphics[width=.6\columnwidth]{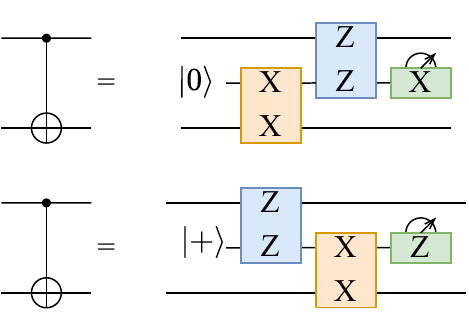}
    \caption{The \textbf{LS} rewrite can decompose a CNOT into any of the two functionally equivalent forms~\cite{horsman2012surface}. A CNOT will require an ancilla, one XX and one ZZ two-body interaction. The figure does not include the corrective terms.}
    \label{fig:ls}
\end{figure}

We start by decomposing an arbitrary Pauli string $P$ into a string $C = C_0 \otimes \ldots \otimes C_n$ of single qubit Clifford gates $C_i$, and a string $P_z = Z^{\otimes n}$, such that $P = C P_z C^\dagger$. As an example, single qubit Clifford gates can be efficiently implemented in practice~\cite{geher2024error, gidney2024inplace} in surface codes. We continue by decomposing $P_z$ via the \textbf{PG} rule~\cite{cowtan2019phase} (Fig.~\ref{fig:pg}). CNOTs are expressed in lattice surgery style~\cite{horsman2012surface} decompositions (Fig.~\ref{fig:ls}). After using \textbf{PG}, we choose the first \textbf{LS} rewrite~\cite{horsman2012surface} for the lhs. CNOT, and the second rewrite for the rhs. CNOT.

\begin{figure}[!h]
    \centering
    \includegraphics[width=.9\columnwidth]{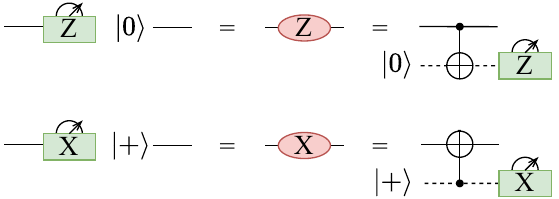}
    \caption{The \textbf{MR} rewrite is joining a measurement and a reset in the same basis into an operation represented by the oval element (e.g. $Z$). Instead of explicitly measuring and resetting, we will consider the circuits on the rhs for achieving this task: the dotted ancilla is initialized, entangled and measured to achieve the measure and reset effect. The figure does not include the corrective terms, but these can be found in the Supplementary Material.}
    \label{fig:mr}
\end{figure}

We replace a sequence of measurement and reset operations in the same basis with an equivalent measurement-based implementation by using the \textbf{MR} rule (Fig.~\ref{fig:mr}). This is, to the best of our knowledge, a novel quantum circuit rewrite rule.

\begin{figure}[!h]
    \centering
    \includegraphics[width=.5\columnwidth]{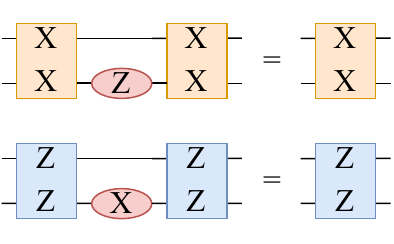}
    \caption{The \textbf{FUSE} reduces the depth of the decomposition by absorbing an oval X or Z element.}
    \label{fig:fuse}
\end{figure}

The \textbf{MR} rewrite rule is applicable for the optimization of circuits which benefit from qubit recycling schemes~\cite{paler2016wire, jiang2024qubit}. By merging the measurement and the reset, it is possible to commute gates across seemingly disjoint parts of the circuit. This has the potential to generate novel optimization heuristics. In particular, this approach has enabled us to derive the \textbf{FUSE} rule (Fig.~\ref{fig:fuse}): we can merge two two-body interactions into a single one, if a measure-and-reset acts on one of the qubits between them.

\subsection{Generalization}

So far, we considered constant depth decompositions of Pauli strings of the form $Z^{\otimes n}$. Nevertheless, there might be qubits which are not involved in the operation, and the corresponding Pauli string will include an $I$ term for those qubits. 

Without loss of generality, we discuss the constant depth, nearest-neighbour implementation of exponentials of the form $Z\otimes \ldots I \ldots \otimes Z$ -- a single $Z$ at the start and the end of the string and at least one $I$ in between. 

Considering the circuit from Fig.~\ref{fig:decomp}, the challenge is to \emph{jump} over the qubits which should not be operated by $Z$ operators -- these qubits are left unchanged by the $I$. In the following, we consider the position (index) of the $Z$ operators in the string (\emph{first}, \emph{last}), such that the discussion will revolve around storage and layouts.

One solution is to permute the qubits such that all $I$ terms are moved to the end on the Pauli string. For example, $Z\otimes \ldots I \ldots \otimes Z$ will be implemented via a single two-body nearest-neighbour interaction $Z\otimes Z \ldots \otimes I$. However, sorting introduces two SWAP networks (one before the Pauli string exponential, and another one afterwards) of logarithmic depth (e.g.~\cite{cowtan2019phase}) instead of a constant depth.

Another solution is to use qudits (moving the qubit state into a higher dimension in order to protect it from qubit $Z$ gates) or special layouts which enable the constant depth construction of Bell pairs using the method from~\cite{beverland2022surface}.

\subsection{Qudits}

It is possible, but not necessarily practical with the current generation of quantum computers, to embed the qubit computation into a system of qudits. To this end, we choose ququarts whose states are spanned by the following basis states: $\ket{0}$, $\ket{1}$, $\ket{2}$, $\ket{3}$. We will assume that all the wires are ququarts instead of qubits.

We make the observation that qubit gates can be applied on qudits~\cite{lanyon2009simplifying} and that whenever this happens, the higher levels of the ququart are left unchanged. For example, the qubit $Z_2$ gate (the subscript indicates the dimension of the qudit, e.g. two is for qubits, and four for ququarts) can be embedded into a $4 \times 4$ matrix written in block form as 
$Z_4=\big(\begin{smallmatrix}
  Z_2 & 0\\
  0 & I_2
\end{smallmatrix} \big)$.

At this stage, whenever we want to apply a qubit operator $I_2$, we can now move the qubit state two levels higher with the ququart $X_4$ gate~\cite{wang2020qudits} which acts as $X_4\ket{i} = \ket{(i+1)\%4}$, such that $X_4^2(a\ket{0}+b\ket{1})= a\ket{2}+b\ket{3}$. By assuming $I_2$ embedded into $I_4$, we can write the following (as our $Z_4$ gate leaves the higher states unchanged):
\begin{align*}
I_2(a\ket{0}+b\ket{1})=X_4^2Z_4X_4^2(a\ket{0}+b\ket{1})
\end{align*}

Therefore, whenever ququarts and the $X_4$ and $Z_4$ gates are supported, we can implement the qubit $Z\otimes \ldots I \ldots \otimes Z$ Pauli string by replacing each $I$ with the single ququart $X_4^2 Z_4 X_4^2$. The result is a continuous string of qubit $Z$ operators which can be implemented according to the decomposition from Fig.~\ref{fig:decomp}.

\begin{figure}[!h]
    \centering
    \includegraphics[width=.4\columnwidth]{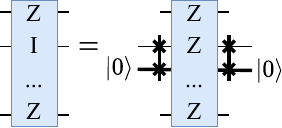}
    \caption{A next-nearest neighbour setup in order to implement the Pauli string in constant depth: using an ancilla initialized in $\ket{0}$ and swapped before and after the Pauli string. The ancilla is left in the $\ket{0}$ state. The "=" sign denotes logical equivalence, as the Hilbert spaces of the two circuits are not equivalent.}
    \label{fig:quq}
\end{figure}

\begin{figure*} 
    \centering
    \includegraphics[width=\textwidth]{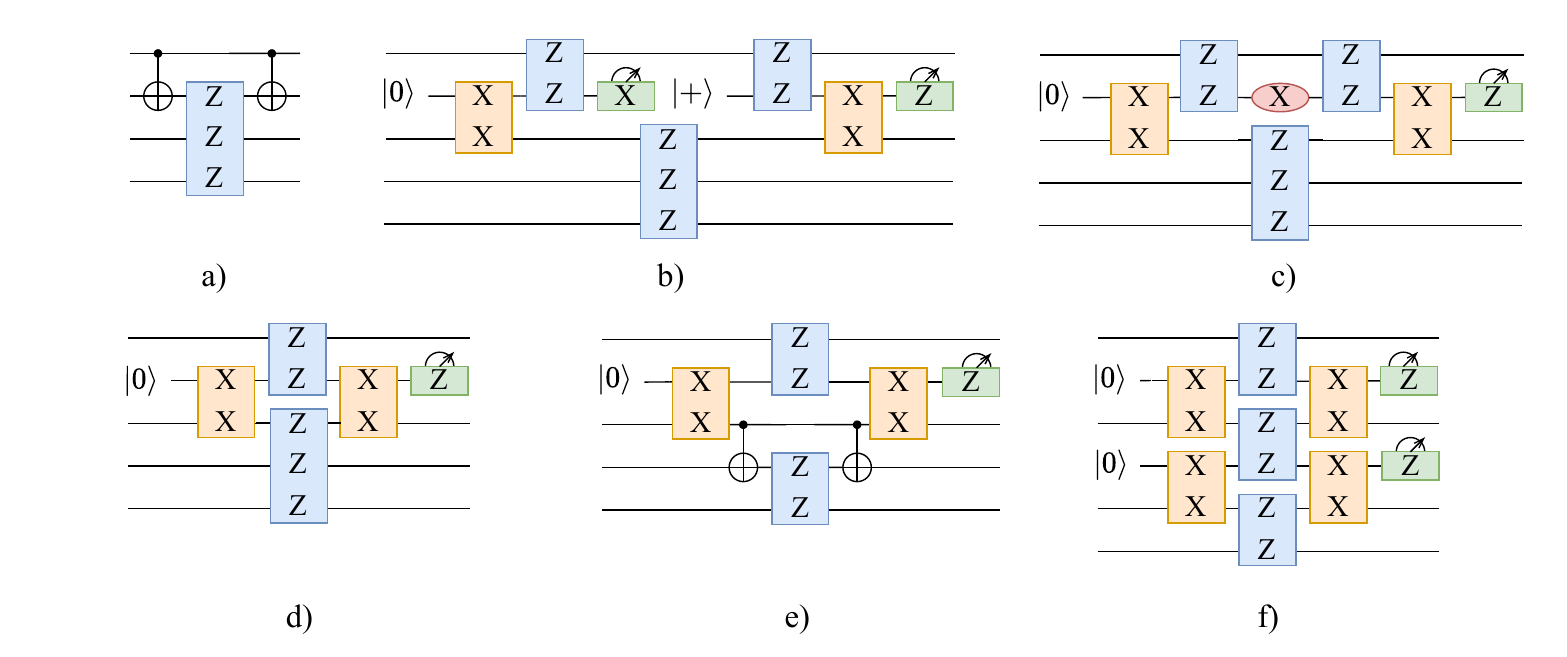}
    \caption{Correctness of the generalized decomposition, illustrated by an example of decomposing a Pauli string of four Z terms: a) applying the \textbf{PG} rule; b) decomposing the two CNOTs using \textbf{LS}; c) the X-measurement and the initialization in $\ket{+}$ can be merged using \textbf{MR} into a red oval X element; d) the two ZZ parities surrounding the red oval are fused using \textbf{FUSE}; e) \textbf{PG} is applied again and inserts two CNOTs; f) the CNOTs are decomposed with \textbf{LS}, the measure and reset are merged with \textbf{MR} and, finally, \textbf{FUSE} is applied.}
    \label{fig:appb}
\end{figure*}

If ququarts are not an option, qubits can be used to simulate the effect of ququarts. Using Fig.~\ref{fig:appc}a, one can show that $e^{i\theta P_1}\ket{A\ldots B\ldots C} = e^{i\theta P_2}\ket{A\ldots 0\ldots C}$ for $P_1=Z \ldots I \ldots Z$ and $P_2 = Z \ldots Z \ldots Z$. This means that, whenever the Pauli string includes an $I$ on a particular qubit (e.g. $B$) then it is possible to \emph{swap} $B$ with a $\ket{0}$ and to have a Pauli string that includes a $Z$ term instead of $I$. Therefore, an alternative to using ququarts is to use one additional qubit $q$ placed next to each ancilla $a$ from Fig.~\ref{fig:decomp}. These $q$ qubits will be initialized in $\ket{0}$, and the $I$ operators from the Pauli string would be replaced by $SWAP(a,q)$, a $Z$ on $a$, followed by $SWAP(a,q)$ (Fig.~\ref{fig:quq}). This construction would not be nearest-neighbour, but next-nearest neighbour.

\subsection{The qubit layout}

The nearest-neighbour implementation of the two-body interaction is only possible if the qubit layout supports it. For example, in Fig.~\ref{fig:lsl} it is not possible to implement the blue merge operations between nearest-neighbour patches, because each logical Q-patch is surrounded by one ancilla A-patch and three routing patches (not marked by any letter).

In a \emph{linear} nearest neighbour layout of qubits, the logical qubits are arranged in a line similar to $\begin{smallmatrix}
  Q & 0 & Q & 0 \ldots & Q
\end{smallmatrix}$, where $0$ are ancillae initialized in $\ket{0}$. In this scenario, pairwise nearest-neighbour interactions are not possible without resorting to ququarts, as discussed in the previous section. Consequently, the long range construction from~\cite{baumer2024measurement} uses nearest-neighbour pairwise interactions, but it is limiting the state of the intermediate qubits to the less general layout $\begin{smallmatrix}
  Q & 0 & 0 & 0 \ldots & Q
\end{smallmatrix}$ --  all the qubits between the first and last one are initialized in $\ket{0}$.

Yet another solution to achieving constant depth implementation of the Pauli string is to use a \emph{bi-linear} nearest neighbour layout of qubits of the form $\begin{smallmatrix}
  Q & 0 & Q & 0 \ldots & Q\\
  r & r & r & r \ldots & r
\end{smallmatrix}$. This layout allows for \emph{jumping} over qubits operated by $I$ and maintaining the nearest-neighbour decomposition. If the $r$ qubits would be operated in a fault-tolerant manner, protected by surface codes, these can be used to first prepare Bell pairs using the constant depth recipe from Fig.~18 in~\cite{beverland2022surface} and then to follow the Pauli string implementation using Bell pairs~\cite{beverland2022surface, yang2024harnessing, buhrman2024state}. Simultaneously, the layout from Fig.~\ref{fig:lsl} has the same properties the bi-linear nearest neighbour layout has.

\section*{Data Availability}
No data was generated or used for this work.

\section*{Code Availability}
No code was generated or used for this work.

\section*{Acknowledgments}
We thank Alexander Cowtan, Matthew Steinberg, M. Sohaib Alam, Gy\"orgy Geh\'er for the stimulating discussions,  Arshpreet Singh Maan and Huyen Do for their feedback, Ryan Babbush for suggesting the application to Adapt-VQE and the Centro de Ciencias de Benasque Pedro Pascual for hosting us while preparing the manuscript. This research was developed in part with funding from the Defense Advanced Research Projects Agency [under the Quantum Benchmarking (QB) program under award no. HR00112230006 and HR001121S0026 contracts], and was supported by the QuantERA grant EQUIP through the Academy of Finland, decision number 352188. The views, opinions and/or findings expressed are those of the author(s) and should not be interpreted as representing the official views or policies of the Department of Defense or the U.S. Government.

\section*{Author Contributions}
I.M. and A.P. conceived the method, I.M. and A.P. wrote the manuscript, I.M. and A.P. contributed equally to the verification of the results.

\section*{Competing Interests}
The authors declare no competing interests.

\section*{Corresponding Authors}
Correspondence to Ioana Moflic (\url{ioana.moflic@aalto.fi}) or Alexandru Paler (\url{alexandru.paler@aalto.fi}).

% \appendix

\begin{figure*}
\centering
    \includegraphics[width=\textwidth]{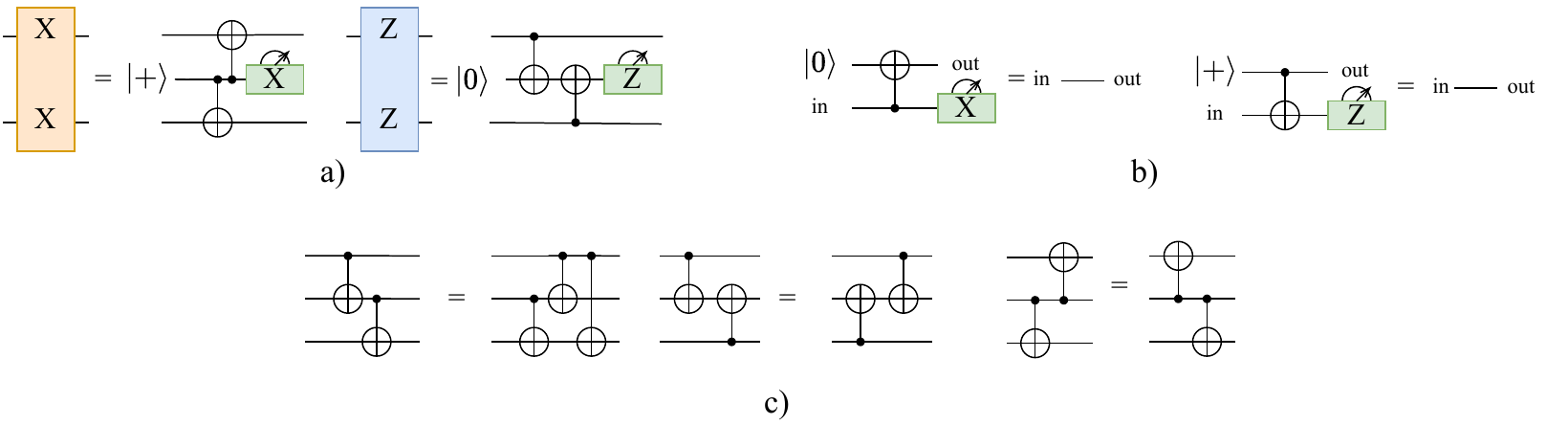}
    \caption{Rewrite rules used as building blocks for more complex circuit identities (the figure does not include the corrective terms): a) the \textbf{XXC} and \textbf{ZZC} are circuit decompositions for the two-body XX and ZZ interactions; b) \textbf{REMZ} and \textbf{REMX} are replacing teleportation-like sub-circuits with the identity; c) \textbf{CNOT} commutation rules.}
    \label{fig:appc}
\end{figure*}

\begin{figure*}
    \centering
    \includegraphics[width=\textwidth]{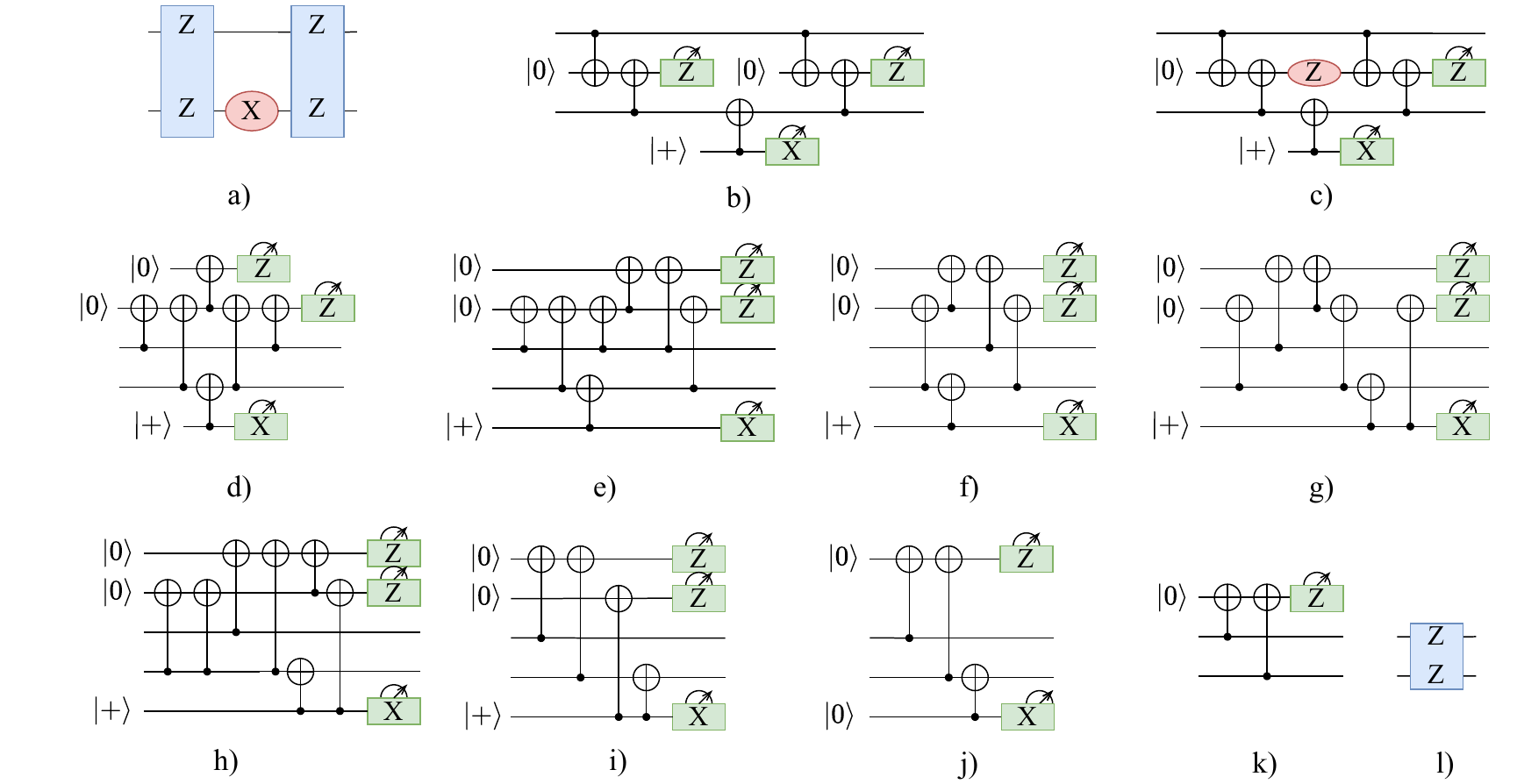}
    \caption{Correctness of the \textbf{FUSE} rewrite rule uses the rewrites listed in Fig.~\ref{fig:appc}  (the figure does not include the corrective terms): a) the start configuration; b) each ZZ interaction is decomposed using \textbf{ZZC} and the red X oval is decomposed using the circuit from \textbf{MR}; c) applying \textbf{MR} between the Z-measurement and the $\ket{0}$ initialization introduces a red Z oval; d) the Z oval is decomposed using the circuit from \textbf{MR}; e-i) CNOTs are commuted according to the \textbf{CNOT} rule, adjacent CNOTs are cancelled; i) the second ancilla from the top is removed by applying \textbf{REMX}; j) the CNOT is controlled by 0, therefore it has no effect and can be safely removed; k) \textbf{ZZC} is applied in the reverse direction; l) end result.}
    \label{fig:appa}
\end{figure*}

\section*{References}
\bibliographystyle{naturemag}
\bibliography{__main}

\clearpage
\appendix*

\renewcommand{\figurename}{Supplementary Figure.}
\setcounter{figure}{0}
\renewcommand{\theequation}{\arabic{equation}}
\setcounter{equation}{0}

\section*{Supplementary Material: Circuit rewrite proofs} 
\label{sec:appendix}

We present mathematical proofs for the circuit identities \textbf{ROT}, \textbf{MR}, \textbf{REM}, and \textbf{FUSE}, as well as additional explanations on the corrective terms omitted in the figures of the main text.

In general, Pauli exponentials can be obtained by conjugating $e^{-i\theta P}$ with Clifford gates as $C(e^{-i\theta P})C^\dagger = e^{-i\theta (C P C^\dagger)}$. In the following, we will assume $P = Z^{\otimes n}$ for simplicity. Throughout the rest of the manuscript, we will use the general definition of projectors $\Pi_{\pm} = \frac{1}{2}(I \pm O)$ to denote the action of the measuring observables $O$ with eigenvalues $\pm1$. By convention, the subscript $q$ of the single-qubit operators $O_q$ will refer to the qubit index the operator acts on. In the case of two-qubit operators $O_{c,t}$, the subscripts indicate the control and target qubit index, respectively. At some point we will use $P^c$ to denote a corrective Pauli operator $P$.

\subsection{\textbf{ROT} rule proof}
\label{appendix:rot}
We show that the circuit from Fig.~\ref{fig:litinski} implements the $P(\theta) = e^{-i\theta/2 P} = \cos(\theta/2)I - i\sin(\theta/2)P$ operator on the data qubits by extending the Pauli string with a Z term on an ancilla initialized in $ \ket{\theta} = R_z(\theta)\ket{+} = \frac{1}{\sqrt{2}}(\ket{0} + e^{i\theta}\ket{1})$ (we use the definition of $R_z$ that includes a global phase of $e^{-i\theta/2}$) and measured in the X basis at the end. The n-qubit system $\ket{\psi}$, together with the ancilla $\ket{\theta}$, starts in the state:
\begin{align}
\ket{\Psi} = \ket{\psi} \ket{\theta} = \frac{1}{\sqrt{2}} (\ket{\psi}\ket{0} + e^{i\theta}\ket{\psi}\ket{1})
\end{align}

The measurement projectors are 
\begin{align}
\Pi_{\pm} = \frac{1}{2}(I \pm P \otimes Z) = \frac{1}{2}(I + mP \otimes Z)
\end{align}
, where $m$ the sign of the measurement outcome of the $P \otimes Z$ observable. The application of the projector results in $\Pi_{\pm} \ket{\Psi}$:
\begin{align}
    % \Pi_{\pm} \ket{\Psi} 
    &\frac{1}{2\sqrt{2}} (I + mP \otimes Z)(\ket{\psi}\ket{0} + e^{i\theta}\ket{\psi}\ket{1})\\
    &= \frac{1}{2\sqrt{2}} (\ket{\psi}\ket{0} + e^{i\theta} \ket{\psi}\ket{1} + mP\ket{\psi}\ket{0} - me^{i\theta}P\ket{\psi}\ket{1}) \nonumber\\
    &= \frac{1}{2\sqrt{2}} \big[ \ket{\psi} \otimes (\ket{0} + e^{i\theta}\ket{1}) +  mP\ket{\psi} \otimes (\ket{0} - e^{i\theta}\ket{1}) \big].\nonumber
\end{align}

By using 
\begin{align}
\ket{0} = \frac{1}{\sqrt{2}}({\ket{+} + \ket{-}}), \ket{1} = \frac{1}{\sqrt{2}}({\ket{+} - \ket{-}})
\end{align}
we substitute $\ket{0} \pm e^{i\theta}\ket{1}$ with 
\begin{align}
    \frac{1}{\sqrt{2}}\big[ (1 \pm e^{i\theta})\ket{+} + (1 \mp e^{i\theta})\ket{-}\big]
\end{align}
and obtain $\Pi_{\pm} \ket{\Psi}$:
\begin{align}
\frac{1}{4}\biggr[ \ket{\psi} \otimes \big[ (1 + e^{i\theta})\ket{+} + (1 - e^{i\theta})\ket{-}\big] + \\
mP \ket{\psi} \otimes \big[ (1 - e^{i\theta})\ket{+} + (1 + e^{i\theta})\ket{-}\biggr].
\end{align}

After measuring the ancilla in the X basis, we obtain two states corresponding to the $\pm1$ eigenvalues which we are rewriting by using trigonometric identities relating full angles to half angles.
\begin{align}
    \ket{\Psi_+} &= \frac{1}{4}\biggr[(1 + e^{i\theta})\ket{\psi} + (1 - e^{i\theta})P\ket{\psi}\biggr] \\
    &= \frac{1}{2} e^{i\theta/2} \biggr[\cos\left(\theta/2 \right)\ket{\psi} - i \sin\left(\theta/2 \right)P\ket{\psi} \biggr] \nonumber\\
    &= \frac{1}{2} e^{i\theta/2} \biggr[\cos\left( \theta/2 \right)I - i \sin\left( \theta/2 \right)P \biggr]\otimes \ket{\psi}\nonumber\\
    &\propto e^{-i\theta/2 P}\ket{\psi}\nonumber\\
    \ket{\Psi_-} &= \frac{1}{4} \left[ (1 - e^{i\theta})\ket{\psi} - (1 + e^{i\theta})P\ket{\psi} \right] \\
    &= \frac{1}{2} e^{i\theta/2} \biggr[-\cos \left( \theta/2 \right) P\ket{\psi} - i\sin\left( \theta/2 \right)\ket{\psi} \biggr] \nonumber\\
    &= -\frac{1}{2} e^{i\theta/2} P \biggr[\cos\left( \theta/2 \right)I + i\sin\left( \theta/2 \right)P\biggr]\otimes \ket{\psi}\nonumber\\
    &\propto P^c e^{i\theta/2 P}\ket{\psi}\nonumber
\end{align}

Thus, up to global phase, we have the two possible states after measuring the ancilla, which are 
\begin{align}
\ket{\Psi_+} &= e^{-i\theta/2 P}\ket{\psi}\\
\ket{\Psi_-} &= P^c e^{i\theta/2 P}\ket{\psi} = e^{i\theta P}P^c\ket{\Psi_+}
\end{align}
The latter state requires: 1) a Pauli correction $P^c$ which, because it is a Clifford gate, can be tracked in classical software and 2) an additional rotation by the double of the angle $\theta$ in order to correct the negative angle (i.e. $R_z(\theta) = R_z(2\theta)R_z(-\theta)$).

% will be tracked and corrected in the Pauli frame through classical software depending on whether the X-basis measurement of the ancilla has outcome $-1$ ($P(-2\theta)$ tracked and corrected) or $+1$ (no correction applied).

\subsection{\textbf{PG} rule proof}

The \textbf{PG} rule (Fig.~\ref{fig:pg}) is shown by induction on the weight $n$ of the Pauli string. We start with the base case $n=1$: there is no reduction applicable, and the minimum weight is achieved.

For the case of $n=2$, we show that for an arbitrary angle $\theta$, $CX_{1,2} e^{i\theta Z_2} CX_{1,2} = e^{i\theta Z_1Z_2}$. Using the circuit identity
\begin{align}
    CX_{c,t} e^{i\theta Z_t} CX_{c,t} = e^{i\theta Z_cZ_t}
\end{align} 
we can show that
\begin{align}
    &CX_{1,2} e^{i\theta Z_2} CX_{1,2} \\
    &= \cos\left(\theta\right) CX_{1,2} I CX_{1,2} + i\sin\left(\theta\right) CX_{1,2} Z_2 CX_{1,2}\nonumber\\
    &= \cos\left(\theta\right) I + i\sin\left(\theta\right) Z_1 Z_2 \nonumber\\
    &= e^{i\theta Z_1Z_2}\nonumber
\end{align}

% Similarly, for $n=3$, we get $CX_{1,2}e^{i\theta Z_2Z_3}CX_{1,2} = CX_{1,2} \biggr[CX_{2,3} e^{i\theta Z_3} CX_{2,3}\biggr]CX_{1,2}$ = .
We proceed with the induction hypothesis, and assume that the recursive decomposition of Fig.~\ref{fig:pg} holds for $n=k$, and have: 
\begin{align}
    e^{i\theta Z_1 \otimes Z_2 \otimes \ldots\otimes Z_k} = CX_{k, k-1}\ldots CX_{2,1}e^{i\theta Z_1}CX_{2,1}\ldots CX_{k, k-1}
\end{align}

Induction Step ($k \rightarrow k+1$). We show 
\begin{align}
    e^{i\theta Z^{\otimes k+1}} = CX_{k+1, k} e^{i\theta Z^{\otimes k}} CX_{k+1, k}
\end{align}
using the induction hypothesis,
\begin{align}
    & CX_{k+1, k} e^{i\theta Z^{\otimes k}} CX_{k+1, k} \\
    &= CX_{k+1, k} \biggr[\ldots CX_{2,1} e^{i\theta Z_1}CX_{2,1} \ldots \biggr]CX_{k+1, k} \nonumber\\
    &= \cos(\theta) I + i\sin(\theta)Z_1Z_2\ldots Z_kZ_{k+1} \nonumber\\
    &= e^{i\theta Z_1Z_2\ldots Z_kZ_{k+1}}.\nonumber
\end{align}
The decomposition from Fig.~\ref{fig:pg} holds for $n=k+1$, and by induction, it holds for arbitrarily large $n$.

\subsection{\textbf{LS} rule proof}

The LS rule represents the lattice-surgery implementation of the logical CNOT gate. Correctness proofs of this circuit rewrite are presented in multiple works~\cite{horsman2012surface, vuillot2019code, de2020zx, erhard2021entangling}. In Supplementary Fig.~\ref{fig:ls_corrections}, we include the correction terms which are missing in Fig.~\ref{fig:ls}.
\begin{figure}[!h]
    \centering
    \includegraphics[width=0.9\columnwidth]{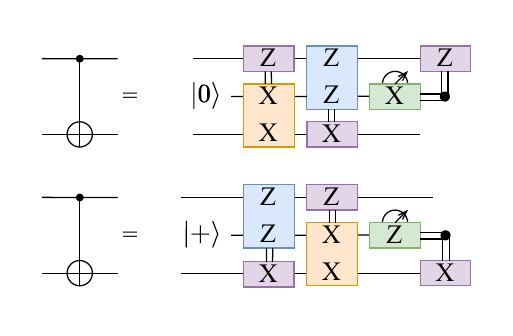}
    \caption{The \textbf{LS} rule with corrective terms included.}
    \label{fig:ls_corrections}
\end{figure}

\subsection{\textbf{MR} rule proof}
\begin{figure}[!h]
    \centering
    \includegraphics[width=0.9\columnwidth]{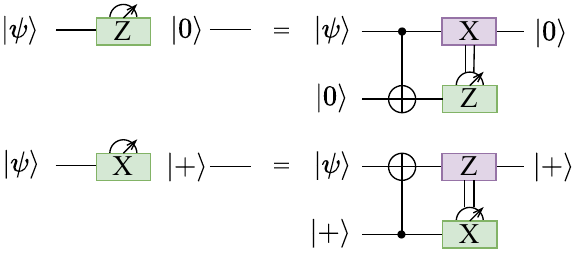}
    \caption{The \textbf{MR} rule with corrective terms included.}
    \label{fig:mr_corrections}
\end{figure}

The \textbf{MR} rule of Fig.~\ref{fig:mr} models an identity operation by joining wires which are measured and a reset in the same basis. We initialize an ancilla in the $+1$ eigenstate of the measurement observable (this is the value we reset to) and teleport the value to the original qubit by entangling it to the ancilla and measuring the observable. To ensure the reset in the $+1$ eigenstate, we will apply a Pauli correction (Supplementary Fig.~\ref{fig:mr_corrections}). 

% mai e nevoie sa explic tot?...
\subsubsection{MRZ rule proof}
The system starts in an unknown state 
\begin{align}
    \ket{\psi}\ket{0} = a \ket{00} + b \ket{10}
\end{align}
($a$, $b$ $\in \mathbb{C}$ and $|a|^2 + |b|^2 = 1$) and becomes 
\begin{align}
    a \ket{00} + b \ket{11}
\end{align} 
after the CX. The measurement projectors are 
\begin{align}
    \Pi_\pm = \frac{1}{2}(II \pm IZ_2) = \frac{1}{2} I \otimes (I \pm Z_2)
\end{align}

After measuring the ancilla in the Z basis, we have two outcomes for the original qubit: $+1$ measurement outcome corresponds to state $\ket{0}$ and the $-1$ measurement outcome corresponds to state $\ket{1}$. For the latter, a Pauli-X correction is required to reinitialize in the qubit in the correct state.

\subsubsection{MRX rule proof}
Similarly to \textbf{MRZ}, we end up with system state $\frac{1}{\sqrt{2}}(\ket{\psi}\ket{0} + X_1\ket{\psi}\ket{1})$ right before measurement. The measurement projectors are $\Pi_\pm= \frac{1}{2} I \otimes (I \pm X_2)$ and the post-measurement state of the first qubit is either $\ket{+}$ or $\ket{-}$, up to a global phase. Again for the latter case, a Pauli-Z correction will reset in the correct basis. 

\subsection{XXC and ZZC rules proof}

The circuit identities of Fig.~\ref{fig:appc}(a) show the implementation of XX and ZZ multi-body measurements via CNOT gates. We will show that the right-hand side in both circuits implements projectors $\Pi_\pm^{X} = \frac{1}{2}(II \pm XX)$ and $\Pi_\pm^{Z} = \frac{1}{2}(II \pm ZZ)$, respectively. The corrections required to enforce the projection into the $+1$ eigenspace are in Supplementary Fig.~\ref{fig:xxczzc_corrections}.

\begin{figure}[!h]
    \centering
    \includegraphics[width=0.6\columnwidth]{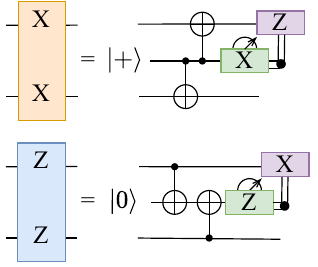}
    \caption{The \textbf{XXC} and \textbf{ZZC} rules are the original circuits from Fig.~
    \ref{fig:appc} and include the additional corrective terms.}
    \label{fig:xxczzc_corrections}
\end{figure}

\subsubsection{XXC rule proof}
We start with a three-qubit system state \begin{align}\ket{\Psi} = \ket{\psi}\ket{+}\end{align}, where \begin{align}\ket{\psi} = a\ket{00} + b\ket{01} + c\ket{10} + d\ket{11}\end{align} with $a$, $b$, $c$, $d \in \mathbb{C}$ and $|a|^2 + |b|^2 + |c|^2 + |d|^2 =1$.

After the two CNOT gates, we are left with \begin{align}\ket{\Psi} = \frac{1}{\sqrt{2}} (\ket{\psi}\ket{0} + XX\ket{\psi}\ket{1})\end{align}. 

Measuring the ancilla in the X basis will give \begin{align}\ket{\Psi_{\pm}} = \frac{1}{2}(\ket{\psi} \pm XX\ket{\psi})\ket{\pm}\end{align} which implements the projectors $\Pi_\pm^{X}$ of the XX operator. To enforce the projection into the $+1$ eigenspace, we require a Pauli-Z correction on either of the qubits. We get 
\begin{align}
\ket{\Psi_{+}} &= \frac{1}{2}(II+XX)\ket{\psi} \propto \ket{++} +\ket{--}
\end{align}
which is already in the correct eigenspace. Therefore, no correction is required.

On the other hand, 
\begin{align}
\ket{\Psi_{-}} &= \frac{1}{2}(II-XX)\ket{\psi} \propto \ket{+-} + \ket{-+}
\end{align}
will require a $Z$ correction on either of the two qubits.

\subsubsection{ZZC rule proof}
We start with $\ket{\Psi} = \ket{\psi}\ket{0}$, and after the application of the two CNOT gates, we are left with \begin{align}\ket{\Psi} = (a\ket{00}+b\ket{11})\ket{0} + (b\ket{01}+c\ket{10})\ket{1}\end{align}. Measuring the ancilla in the Z basis will give \begin{align}\ket{\Psi_{+}} \propto \ket{00}+\ket{11}\end{align} and \begin{align}\ket{\Psi_{-}} \propto \ket{01}+\ket{10}\end{align} which correctly implement the projectors $\Pi_\pm^{Z}$ of the ZZ operator. Projecting into the $+1$ eigenspace will again require a Pauli-X correction on either of the qubits of $\ket{\Psi_{-}}$, as seen in Supplementary Fig.~\ref{fig:xxczzc_corrections}.

\subsection{REMZ and REMX rules proof}
The \textbf{REMZ} and \textbf{REMX} rules are almost identical to \textbf{MRZ} and \textbf{MRX}, respectively. The only difference is that instead of teleporting the $+1$ eigenvectors of the measurement operator, we are teleporting the state $\ket{\psi}$. In both cases, we will require a correction, just as in Supplementary Fig.~\ref{fig:mr_corrections}.

\begin{figure}[!h]
    \centering
    \includegraphics[width=0.9\columnwidth]{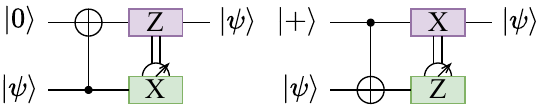}
    \caption{The \textbf{REMX} and \textbf{REMZ} rules with corrective terms included.}
    \label{fig:remxremz_corrections}
\end{figure}

\begin{figure*}
        \centering
    \begin{subfigure}{0.49\linewidth}
        \centering
        \includegraphics[height=4cm]{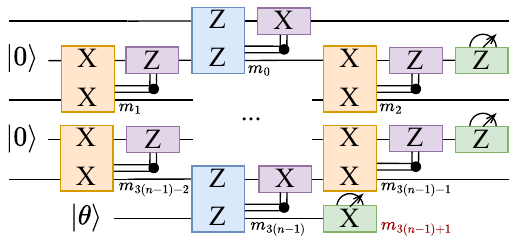}
        \caption{ }
        \label{fig:decomp_2}
    \end{subfigure}
    \hfill
    \begin{subfigure}{0.49\linewidth}
        \centering
        \includegraphics[height=4cm]{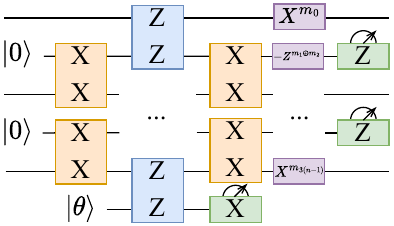}
        \caption{ }
        \label{fig:decomp_3}
    \end{subfigure}
    \caption{(a) The decomposition from Fig.~\ref{fig:decomp} with Pauli corrections added: purple boxes represent the necessary correction terms. Each XX or ZZ operator will give a measurement outcome $m_k$ (depicted in the lower right side of each operator). When $m_k = -1$, a single-qubit Pauli correction is applied. The outcome $m_3(n-1)+1$ (highlighted in red in the figure) of the X-measurement is recorded for applying the final Pauli correction in Supplementary Material Section~\ref{appendix:rot}; (b) The corrections from Fig.~\ref{fig:decomp_2} can be commuted to the end of the circuit. These corrections will be applied only virtually, therefore these do not change the structure of the original decomposition.}
    \label{fig:final_corrections}
\end{figure*}

\subsection{FUSE rule proof}
The \textbf{FUSE} rule (Fig.~\ref{fig:fuse}) is built out of previously proven building blocks such as \textbf{ZZC} and \textbf{MRX}. Additionally, we present a mathematical proof for the correctness proof of Fig.~\ref{fig:appa}. 

We start in an arbtrary two-qubit state \begin{align}\ket{\psi} = a\ket{00} + b\ket{01} + c\ket{10} + d\ket{11}\end{align} with $a$, $b$, $c$, $d \in \mathbb{C}$ and $|a|^2 + |b|^2 + |c|^2 + |d|^2 =1$.

First, we measure the first ZZ operator will yield the states corresponding to the two measurement outcomes $\ket{\Psi_\pm} = \Pi^Z_\pm\ket{\psi} = \frac{1}{2}(II\pm ZZ)(a\ket{00} + b\ket{01} + c\ket{10} + d\ket{11})$. After measurement, we have the normalized states:
\begin{align}
\ket{\Psi^{(1)}_+} = \frac{a\ket{00} + d\ket{11}}{\sqrt{|a|^2 + |d|^2}}, \ket{\Psi^{(1)}_-} = \frac{b\ket{01} + c\ket{10}}{\sqrt{|b|^2 + |c|^2}}
\end{align}

At this stage, a Pauli-X correction will be realized to enforce the projection to the $+1$ eigenspace. Assume no correction is needed and, without loss of generality, we will continue with state $\ket{\Psi^{(1)}_+}$. We use the decomposition of the \textbf{MRX} gate, and add an ancilla initialized in $\ket{+}$ and apply the CNOT gate. We are left with the state
\begin{align}
\frac{1}{\sqrt{2(|a|^2 + |d|^2)}}\biggr[(a\ket{00} + d\ket{11})\ket{0} + (a\ket{01} + d\ket{10})\ket{1}\biggr]
\end{align}

Second, we measure the ancilla in the X basis, we are left with two normalized states, and to proceed we will assume again, without loss of generality, that no correction is needed.
\begin{align}
\ket{\Psi^{(2)}_+} = \frac{a\ket{00} + d\ket{11} +a\ket{01} + d\ket{10}}{\sqrt{2(|a|^2 + |d|^2)}}\\
\ket{\Psi^{(2)}_-} = \frac{a\ket{00} + d\ket{11} - a\ket{01} - d\ket{10}}{\sqrt{2(|a|^2 + |d|^2)}}
\end{align}

Third, we now need to measure the ZZ operator again on $\ket{\Psi^{(2)}_+}$. We get 
\begin{align}
\ket{\Psi^{(3)}_+} = \frac{a\ket{00} + d\ket{11}}{\sqrt{|a|^2 + |d|^2}}\\
\ket{\Psi^{(3)}_-} = \frac{a\ket{01} + d\ket{10}}{\sqrt{|a|^2 + |d|^2}}
\end{align}

After the third measurement, we have projected back to the correct $\pm$ eigenspaces of a single ZZ operator (corresponding to $\ket{\Psi^{(1)}_\pm}$). This proves the correctness of the \textbf{FUSE} circuit identity.

\section*{Supplementary Material: Corrective terms}

The constant depth decomposition from Fig.~\ref{fig:decomp} requires corrective terms in the error-corrected scenario. Herein, we show how Pauli corrections are derived, commuted towards the end of the circuit and applied in software. Although part of the circuit, these corrections would ultimately only be applied in a classical software called \emph{Pauli frame tracker}~\cite{paler2014software}.

In Supplementary Fig.~\ref{fig:xxczzc_corrections}, each ZZ or XX operator can be implemented in a measurement-based manner. This implementation does require Pauli corrections in the final decomposition (Fig.~\ref{fig:decomp_2}): each ZZ operator requires a Pauli-X correction and each XX operator requires a Pauli-Z correction. The correction terms from Fig.~\ref{fig:decomp_2} do not need to be implemented in hardware. These single-qubit Pauli operators can be virtually commuted through the logical circuit. Note that Z-corrections commute with ZZ operators, as $(Z \otimes I)(Z\otimes Z) = (Z\otimes Z)(Z \otimes I)$ and anticommute with XX operators, since $(Z \otimes I)(X\otimes X) = -(X\otimes X)(Z \otimes I)$. Therefore, each Pauli-X correction will be naively commuted to the end of the circuit, while Pauli-Z corrections will cumulate and gather a sign change, as seen in Fig.~\ref{fig:decomp_3}. Practically, this is a similar approach to the one from~\cite{litinski2019game}.

\end{document}